\documentclass{elsart}
\usepackage[normalem]{ulem}
\usepackage{latexsym,amssymb,amsmath,amssymb,xspace,theorem}
\usepackage{multicol}
\usepackage{t1enc}
\usepackage{epic,eepic}
\usepackage{subfigure}
\usepackage{algorithmic}
\theoremstyle{plain}\theoremheaderfont{\scshape}

\newtheorem{Thm}{\bf Theorem}[section]

\newtheorem{Clm}{Claim}[Thm]

\newtheorem{Prop}[Thm]{\bf Proposition}

{\theorembodyfont{\rmfamily}
 \newtheorem{Def}[Thm]{\bf Definition}
 \newtheorem{Rem}[Thm]{\bf Remark}
 
}
 \newtheorem{Problem}[Thm]{\bf Problem}

\newenvironment{Prf}{{\bf \noindent Proof } }{\hfill$\square$\\}
\newenvironment{PrfClaim}{{\bf Proof }}{{\hfill\tiny{$\blacksquare$\\}}}

\newcommand{\ignore}[1]{}
\newcommand{\ligne}{\vspace{0.5cm}}
\newcommand{\trou}{\vspace{5mm} \noindent}

\newcommand{\cqfd}{\unskip\kern 6pt\penalty 500
\raise -2pt\hbox{\vrule\vbox to 10pt{\hrule width 4pt
\vfill\hrule}\vrule}\par}

\newcommand{\ppm}{$\mathcal P=\{P_1,P_2 \ldots, P_k\}\ $}
\newcommand{\stp}{$\mathcal T=\{T_1,T_2 \ldots, T_k\}\ $}
\newcommand{\np}{$\mathcal T=\{T_1,T_2 \ldots, T_{\frac{n}{2}}\}\ $}

\newcommand{\stprime}{$\mathcal T'=\{T'_1,T'_2 \ldots, T'_k\}\ $}
\newcommand{\cubthree}{ $3$-edge colourable cubic graph \xspace}
\newcommand{\cubsimplethree}{simple $3$-edge colourable cubic graph \xspace}

\newcommand{\cubthreev}{ $3$-edge colourable cubic graph}

\newcommand{\threenc}{$\mathcal T$, $\mathcal T'$ and $\mathcal T^{''}$ \xspace}
\newcommand{\threencsv}{$\mathcal T$, $\mathcal T'$ and $\mathcal T^{''}$}
\newcommand{\Ajoute}[1]{#1}
\newcommand{\Enleve}[1]{}
\input{epsf.sty}

\begin{document}
\begin{frontmatter}

\title{On normal partitions in cubic graphs}

\author{J.L. Fouquet and J.M. Vanherpe}

\address{L.I.F.O., Facult\'e des Sciences, B.P. 6759 \\
Universit\'e d'Orl\'eans, 45067 Orl\'eans Cedex 2, FR}

\begin{abstract}
A {\em normal partition} of the edges of a cubic graph is a
partition into {\em trails}  (no repeated edge) such that each
vertex is the end vertex of exactly one trail of the partition. We
investigate this notion and give some results and problems.

\end{abstract}


\end{frontmatter}

\section{Introduction and notations}

\begin{figure}[htb]
\centering \epsfsize=0.2 \hsize \noindent
\epsfbox{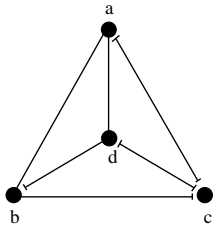} \caption{Normal and eccentric vertices}
\label{Figure:FigureInitiale}
\end{figure}

  Let $G=(V,E)$ be a cubic graph (loops and multiple edges are
allowed) and let \stp be a partition of $E(G)$ into trails (no
repeated edge). Every vertex $v \in V(G)$ is either an end vertex
three times in the partition and we shall say that $v$ is an {\em
eccentric} vertex, or an end vertex exactly once, and we shall say
that $v$ is a {\em normal} vertex. To each vertex $v$ we can
associate a set $E_{\mathcal T}(v)$ containing the end vertices of
the unique trail with $v$ as an internal vertex, when such a trail
exists in $\mathcal T$. When $v$ is eccentric we obviously have
$E_{\mathcal T}(v) = \emptyset$. It must be clear that we can have
$v \in E_{\mathcal T}(v)$ since
 we consider a partition of trails. \Ajoute{In Figure \ref{Figure:FigureInitiale} we have drawn $K_{4}$
with the trail
 partition $\mathcal T =\{bdabc,dc,ac\}$. The vertex $c$ is an
 eccentric vertex while $a$, $b$ and $d$ are normal vertices.}

\begin{Def}\label{Definition:NormalPartition}
A partition \stp  of $E(G)$ into trails is {\em normal} when every
vertex is normal.
\end{Def}
When $\mathcal T$ is a normal partition, we can associate to each
vertex the unique edge with end $v$ which is the end edge of a trail
of $\mathcal T$. We shall denote this edge by $e_{\mathcal T}(v)$
and it will be convenient to say that $e_{\mathcal T}(v)$ is the
{\em marked} edge associated to $v$. When it will be necessary to
illustrate our purpose by a figure the marked edge associated to a
vertex  will be figurate by a $\vdash$ close to this vertex.

\Ajoute{Our purpose, in this paper, is to investigate this new
notion of normal partition. In particular we shall see that normal
odd partitions can be associated in a natural way to perfect
matchings. We shall introduce the notion of {\em compatible normal
partitions} (to be defined later) leading to a property that could
be verified by every bridgeless cubic graph (including the so called
{\em snarks}) and we shall give some results in that direction.}

\begin{Def}\label{Definition:OddPartition}
A partition \stp  of $E(G)$ into trails is {\em odd} when every
trail in $\mathcal T$ is odd.
\end{Def}

\begin{Def}\label{Definition:PathPartition}
A partition \stp  of $E(G)$ where each trail is a path will be
called a {\em path partition}.
\end{Def}

\begin{Def}\label{Definition:PerfectPathMatching}
A partition  \ppm of $V(G)$ into paths is a {\em perfect path
partition} when every vertex of $G$ is contained in $\mathcal P$
(\Ajoute{let us note that $k \leq \frac{n}{2}$}). A perfect matching
is thus a perfect path partition where each path has length $1$.
\end{Def}

\textbf{ Notations:} Following Bondy \cite{Bon1}, a {\em walk} in a
graph $G$ is sequence $W := v_0e_1v_1 \ldots e_kv_k$,
 where $v_0, v_1, \ldots , v_k$ are vertices of $G$, and $e_1, e_2 \ldots , e_k$ are edges
 of $G$ and $v_{i-1}$ and $v_i$ are the ends of $e_i$, $1 \leq i \leq k$. The vertices $v_0$ and $v_k$ are the {\em end
 vertices} and $e_1$ and $e_k$ are the {\em end edges} of this walk, while $v_1, \ldots, v_{k-1}$ are the
 {\em internal vertices} and $e_2, \ldots, e_{k-1}$ are the {\em internal edges}. The {\em length} $l(W)$ of $W$ is
 the number of edges (namely $k$). The walk $W$ is {\em odd} whenever $k$ is odd and {\em even} otherwise.

 The walk  $W$ is a {\em trail} if its edges
 $e_1, e_2, \ldots, e_k$ are \Ajoute{distinct} and a {\em path} if its vertices
 $v_0, v_1, \ldots, v_k$ are \Ajoute{distinct}. If $W := v_0e_1v_1 \ldots e_kv_k$, is a walk of
 $G$, $W':=v_ie_{i+1} \ldots e_jv_j$
 ($0 \leq i \leq j \leq k$) is a {\em subwalk}  of $W$ ({\em  subtrails} and
 {\em subpaths} are defined analogously).

 If $v$ is an internal vertex of a walk $W$ with ends $x$ and $y$,
 $W(x,v)$ and $W(v,y)$ are the subwalks of $W$ obtained \Ajoute{by} cutting $W$ in
 $v$. Conversely if $W_1$ and $W_2$ have a common end $v$, the
 {\em concatenation} of these two walks {\em on $v$} gives rise to a new
 walk (denoted by $W_1+W_2$) with $v$ as an internal vertex.
 When no confusion, is possible, it
 will be convenient to omit the edges in the description of a walk,
 that is $W := v_0e_1v_1 \ldots e_kv_k$ will be shorten in $W := v_0v_1 \ldots
 v_k$.

When $F \subseteq E(G)$, $V(F)$ is the set of vertices which are
incident with some edge of $F$ and $G-F$ is the graph obtained from
$G$ \Ajoute{by} deleting the edges of $F$. A {\em strong matching}
$C$ in a graph $G$ is a matching $C$ such that there is no edge of
$E(G)$ connecting any two edges of $C$, or, equivalently, such that
$C$ is the edge-set of the subgraph of $G$ induced on the vertex-set
$V(C)$.

\section{Elementary properties}

\begin{Prop} \label{Proposition:ExistenceNormal} Let  $G$ be a cubic
graph. \Ajoute{Then} we can find a normal partition of $E(G)$ within
a linear time.
\end{Prop}
\begin{Prf}
We can easily obtain a partition \stp of $E(G)$ into trails via a
greedy algorithm. If  every vertex is normal then $\mathcal T$ is
normal and we are done. If $v$ is an eccentric vertex then $v$ is
the end vertex of two distinct trails $T_1$ and $T_2$. Let $T'$ be
the trail obtained by concatenation of $T_1$ and $T_2$ on $v$.
\Ajoute{Then} $v$ is an internal vertex of $T'$ and
$T-\{T_1,T_2\}+T'$ is a partition of $E(G)$ into trails with one
eccentric vertex less (namely $v$). This operation can be repeated
as long as the current partition into trails has an eccentric vertex
and we end with a normal partition in at most $O(n)$ steps.
\end{Prf}

\begin{Prop} \label{Proposition:NormalNover2}
A partition  $\mathcal T$ of \Ajoute{a cubic graph }$G$ is normal if and only if $|\mathcal
T|=\frac{n}{2}$.
\end{Prop}
\begin{Prf}
Assume that $\mathcal T$ is normal, then every vertex is the end of
exactly one trail. Hence $|\mathcal T|=\frac{n}{2}$.

Conversely let $\mathcal T$ be a partition of the edge set of $G$
into trails. Assume that $|\mathcal T|=\frac{n}{2}$ and $T$ is not
normal. Then, performing the operation described in Proposition
\ref{Proposition:ExistenceNormal}  on eccentric vertices leads to a
normal partition $\mathcal T'$ such that $|\mathcal
T'|<\frac{n}{2}$, since the concatenation of two trails on a vertex
decreases the number of trails in the partition, a contradiction.
\end{Prf}

\Ajoute{We shall denote by $n_{\mathcal T}^{i}$ the number of trails
of length $i$}  and by $\mu(T)$ the \Ajoute{average} length of
trails in a \Enleve{normal} partition \Ajoute{$\mathcal T$}.
\begin{Prop} \label{Proposition:Distribution}
Let  \Ajoute{$\mathcal T$} be a normal partition of a cubic graph
$G$ on $n$ vertices. Then

\begin{itemize}
    \item \Ajoute{$\mu(\mathcal T)=3$}
    \item $\sum_{i=1}^{i=n+1}(3-i)n_{\mathcal T}^{i}=0$
\end{itemize}
\end{Prop}
\begin{Prf}
$\mathcal T$ being normal, we have  $|\mathcal T|=\frac{n}{2}$ by
Proposition \ref{Proposition:NormalNover2}. Since
$|E(G)|=\frac{3n}{2}$ we have obviously  $\mu(\mathcal T)=3$.

We have
$$\sum_{i=1}^{i=n+1}i\times n_{\mathcal T}^{i}=\frac{3n}{2}=3\sum_{i=1}^{i=n+1}n_{\mathcal T}^{i}$$

and hence
$$\sum_{i=1}^{i=n+1}(3-i)n_{\mathcal T}^{i}=0$$

\end{Prf}

The {\em length} of a normal partition $\mathcal T$ (denoted by
$l(T)$) is the length of the longest trail in $\mathcal T$.
\Ajoute{Let us note that, by Proposition
\ref{Proposition:Distribution}, every trail of a normal partition
$\mathcal T$ of $G$ has length $3$ when} \Enleve{$\mu(\mathcal
T)=3$} \Ajoute{ $l(\mathcal T) \leq 3$.}

\begin{Prop} \label{Proposition:Hamiltonian}
A cubic graph $G$ on $n$ vertices has an hamiltonian path if and
only if $G$ has a normal partition $\mathcal T$ such that
$l(\mathcal T)=n+1$
\end{Prop}
\begin{Prf}
Assume that $P=v_1v_2\ldots v_n$ is an hamiltonian path of $G$. We
shall consider that $v_i$ is joined to $v_{i+1}$ by the edge $e_i$
in $P$. Let $w_1$ ($w_n$ respectively) \Ajoute{be} a  vertex
adjacent to $v_1$ ($w_1$ respectively) by the edge $e'_1$ ($e'_n$
respectively) not in $E(P)$( $e'_1\neq e'_n$). Let $T_1$ be the
trail $w_1e'_1v_1e_1v_2e_2 \ldots e_{n-1}v_ne'_nw_n$. $E(G)-T_1$ is
reduced to a matching of size $\frac{n-2}{2}$ and it can be easily
checked that this matching together with $T_1$ is a normal partition
of $G$ of length $n+1$.

Conversely let $\mathcal T$ be a normal partition of $G$ of length
$n+1$ and let $T_1=w_1e_1v_1e_1v_2e_2 \ldots e_{n-1}v_ne_nw_n$ be a
trail of maximum length in $T$. Since the only vertices which can
appear twice in $T_1$ are precisely $w_1$ and $w_n$, $P=v_1v_2\ldots
v_n$ is an hamiltonian path of $G$.
\end{Prf}

\begin{Thm} \label{Theorem:FromPerfectPathPartition}Let $G$ be a cubic graph
having a perfect path partition \ppm. Assume that the ends of $P_i$
are $x_i$ and $y_i$ \Ajoute{for every $i=1 \ldots k$.}  Then $G$ has
a normal partition\Enleve{s} \np  such that $T_i$  is obtained from $P_i$ \Enleve{in} \Ajoute{by}
  adding one edge incident to $x_i$ and one
edge incident to $y_i$ \Ajoute{for every $i=1 \ldots k$}.
\end{Thm}

\begin{Prf}
 The subgraph of $G$ obtained \Ajoute{by} deleting the edges of each $P_i$ is
 a set of disjoint paths. Let us give an arbitrary orientation to these
 paths. We get a normal partition $\mathcal T$ \Ajoute{by} adding the outgoing edge
incident to $x_i$ and to $y_i$ (for every $i=1 \ldots k$), the
remaining edges being a set of trails of length $1$ in $\mathcal T$.
\end{Prf}

Let $l_1,l_2 \ldots l_{\frac{n}{2}}$ be a set of integers ($l_i \geq
1$) such that
$$\sum_{i=1}^{\frac{n}{2}}l_i=\frac{3n}{2}.$$
\Enleve{can we}\Ajoute{Is it possible to} find a normal partition \np where $l(T_i)=l_i$ \Ajoute{for
every $i=1 \ldots \frac{n}{2}$}? \Ajoute{We do not know the complete
answer}, however, when $G$ has an hamiltonian cycle we have the
following result (\Ajoute{an extension of a result of}
\cite{BouFou}):

\begin{Thm} \label{Theorem:UniversalNormalPartition}Let $G$ be a cubic hamiltonian graph.
Let $l_1,l_2 \ldots l_{\frac{n}{2}}$ be a set of integers such that

\begin{itemize}
\item $\sum_{i=1}^{\frac{n}{2}}l_i=\frac{3n}{2}$
\item $l_i \geq 1 \quad l_i \not = 2$  $\forall i=1 \ldots \frac{n}{2}$
\end{itemize}
Then $G$ has a normal partition \np where $l(T_i)=l_i$ for every $i=1 \ldots \frac{n}{2}$
\end{Thm}

\begin{Prf} Let $\lambda_i=l_i-2$ and assume that $\lambda_1 \geq
\lambda_2 \geq \ldots \geq \lambda_{\frac{n}{2}}$. The first $k$
values (for some $k \leq  \frac{n}{2}$) are greater than $1$, and
the remaining values are $-1$, since $l_i \not = 2$ for all $i=1
\ldots \frac{n}{2}$. We have
$$\sum_{i=1}^{k}\lambda_i=\sum_{i=1}^{k}(l_i-2)=\sum_{i=1}^{k}l_i-2k$$
$$\sum_{i=1}^{k}l_i-2k=\sum_{i=1}^{k}l_i-2k + \sum_{j=k+1}^{\frac{n}{2}}l_j-(\frac{n}{2}-k)$$
since $\sum_{i=1}^{k}l_i+
\sum_{j=k+1}^{\frac{n}{2}}l_j=\frac{3n}{2}$ we get that
$$\sum_{i=1}^{k}\lambda_i=n-k$$

Let $C$ be an hamiltonian cycle of $G$, we can thus arrange a set
$\mathcal P$ of vertex disjoint paths  $P_i$ of length $\lambda_i$
($i=1 \ldots k$) along this cycle. $\mathcal P$ is a perfect path
partition and, applying Theorem
\ref{Theorem:FromPerfectPathPartition} we have a normal partition of
$G$ as claimed.
\end{Prf}

\begin{figure}[htb]
\centering \epsfsize=0.6 \hsize \noindent \epsfbox{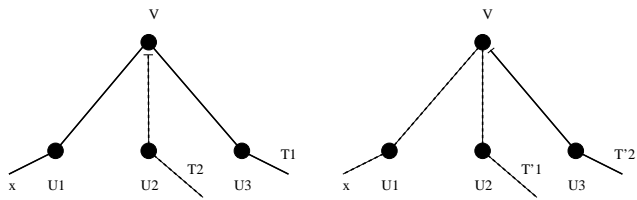}
\caption{Switching on $v$ with two distinct trails}
\label{Figure:SwitchingTwo}
\end{figure}

\begin{figure}[htb]
\centering \epsfsize=0.6 \hsize \noindent \epsfbox{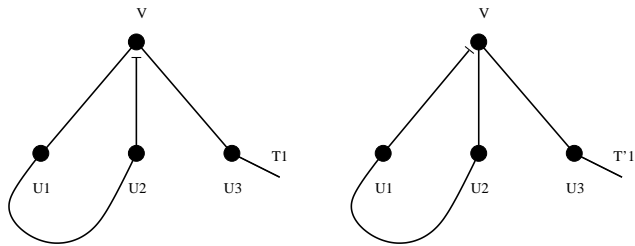}
\caption{Switching on $v$ with one trail}
\label{Figure:SwitchingOne}
\end{figure}

 Let $\mathcal T$ be a normal partition of a cubic graph $G$ and let $v$
be any vertex of $G$. $E_{\mathcal T}(v)$ contains exactly two
vertices, namely $x$ and $y$  and one of them, at least, must be
distinct from $v$ (we may assume that $v \not = x$). Let $T_1$ be
the trail with ends $x$ and $y$ such that $v$ is an internal vertex
of $T_1$. Since $\mathcal T$ is normal, there is a trail $T_2$
ending in $v$ (with the edge $e_{\mathcal T}(v)$). If $T'_1$ denotes
the trail obtained by concatenation of $T_1(x,v)$ and $T_2$ on $v$,
then $\mathcal T-\{T_1,T_2\}+T'_1+T_1(v,y)$ is a new normal
partition of $G$. We shall say that the above operation is a {\em
switch on} $v$. When $v \not \in E_{\mathcal T}(v)$ two such
switchings are allowed (see Figure \ref{Figure:SwitchingTwo}), but
when $v \in E_{\mathcal T}(v)$ only one switching is possible (see
Figure \ref{Figure:SwitchingOne}). A switch on a vertex $v$ (leading
from a normal partition $\mathcal T$ to the normal partition
$\mathcal T'=\mathcal T*v$) does not change the edge marked
associated to $w$ when $w \not = v$. That is $e_{\mathcal
T}(w)=e_{\mathcal T'}(w)$. On the other hand the sets $E_{\mathcal
T'}(w)$ may have changed for vertices of $T_1$ and $T_2$. When
$\mathcal T$ is a normal odd partition and when $\mathcal
T'=\mathcal T*v$ remains \Enleve{to be} an odd partition, the switch
on $v$ is said to be an {\em odd switch}. It is not difficult to see
that, given a normal odd partition, an odd switch is always possible
on every vertex.

\ligne We shall say that $\mathcal T$  and $\mathcal T'$ are  {\em
switching equivalent} (resp. {\em odd switching equivalent})
whenever $\mathcal T'$ can be obtained from $\mathcal T$ by a
sequence of switchings (resp. odd switchings). The {\em switching
class} (resp. {\em odd switching class}) of $\mathcal T$ is the set
of normal partitions which are switching equivalent (resp. odd
switching equivalent) to $\mathcal T$ .

\begin{Thm} \label{Theorem:Switching} Let $G$ be a cubic
graph and let $\mathcal T$ and $\mathcal T'$ be any two normal
(resp. odd) partitions. Then $\mathcal T'$ can be obtained from
$\mathcal T$ by a sequence of (resp. odd) switchings of length at
most $2n$.
\end{Thm}

\begin{Prf} Let $A_{\mathcal T \mathcal T'}=\{v |\ v \in V(G), \  e_{\mathcal T}(v)=e_{\mathcal
T'}(v) \}$ and assume that \Ajoute{$V(G)-A_{\mathcal T \mathcal T'}
\not = \emptyset$} \Ajoute{(otherwise we obviously have  $\mathcal
T=\mathcal T^{'}$)}. We want to pick a vertex in
\Ajoute{$V(G)-A_{\mathcal T \mathcal T'}$} and try to switch the
normal partition $\mathcal T$ \Enleve{in} \Ajoute{on} this vertex
(or $\mathcal T'$) in order to increase the size of
\Ajoute{$A_{\mathcal T \mathcal T'}$} \Ajoute{(formally we have
changed $\mathcal T$ into $\mathcal T_{1}$ and $\mathcal T'$ into
$\mathcal T'_{1}$ and we consider the set $A_{\mathcal T_{1}
\mathcal T'_{1}}$)}. We can suppose that $\mathcal T$ and $\mathcal
T'$ are not  switching equivalent  and, moreover, among the
switching equivalent normal partitions of $\mathcal T$ and those of
$\mathcal T'$, $A_{\mathcal T \mathcal T'}$ has maximum cardinality.

Let $v \not \in A_{\mathcal T \mathcal T'}$ and let $e_1,e_2$ and
$e_3$ be the edges adjacent to $v$. Assume that $e_{\mathcal
T}(v)=e_1$ and $e_{\mathcal T'}(v)=e_2$. Recall that in both
partitions a switch (resp. odd switch) is always possible on $v$.

Consider first a possible switch (resp. odd switch) on $v$ in
$\mathcal T$, \Ajoute{we get hence a new normal partition $\mathcal
T*v$}. If $e_{{\mathcal T}*v}=e_2$ then $A_{\mathcal T*v,\mathcal
T'}=A_{\mathcal T,\mathcal T'}\cup\{v\}$, a contradiction. If by
switching (resp. odd switching) $\mathcal T'$ on $v$ we have
$e_{{\mathcal T'}*v}=e_1$ then $A_{\mathcal T,\mathcal
T'*v}=A_{\mathcal T,\mathcal T'}\cup\{v\}$, a contradiction.
Finally, if $e_{{\mathcal T}*v}\neq e_2$ and $e_{{\mathcal
T'}*v}\neq e_1$ that means that $e_{{\mathcal T}*v}= e_3$ and
$e_{{\mathcal T'}*v}= e_3$, thus $A_{\mathcal T*v,\mathcal
T'*v}=A_{\mathcal T,\mathcal T'}\cup\{v\}$, a contradiction.

Hence any two normal partitions are  switching equivalent (resp. odd
switching equivalent). In order to increase the size of $A_{\mathcal
T \mathcal T'}$, we have seen that we eventually are obliged to
proceed to two switchings on the same vertex (one with $\mathcal T$
and one with $\mathcal T'$). It is \Enleve{thus} clear that we need
at most $2n$ such \Ajoute{switchings} on the road leading to
$\mathcal T'$ from $\mathcal T$.
\end{Prf}

\Enleve{Theorem \ref{Theorem:Switching} suggests a simulated
annealing approach in order to search for a longest path in a cubic
graph. We have got  results in that direction in \cite{DelFouThuVir}
when considering linear partitions (partitions of the edge set of a
cubic graph into two   forests of paths). Instead of using a
switching on a vertex the elementary operation involved was a
switching on an edge, but, in that case, it is not true that any two
linear partitions are switching equivalent.}

\begin{figure}[htb]
\centering \epsfsize=0.2 \hsize \noindent \epsfbox{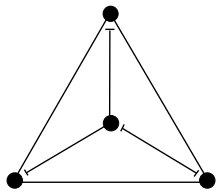}
\caption{No normal partitions associated to the $\vdash$ }
\label{Figure:K4}
\end{figure}

\Ajoute{In Figure \ref{Figure:K4}, we can see that it is not
possible to find a normal partition of $K_{4}$ for which the set of
marked edges is given by those having a $\vdash$ at one end. Since
the set of edges with no end marked contains a cycle the following
question is thus natural.} Given a set of edges $F=\{e_v | v \in
V(G)\}$, where each vertex of $V(G)$ appears exactly once as the end
of an edge of $F$, under which condition can we say that this set of
edges is the set of marked edges associated to a normal partition?

\begin{Thm} \label{Theorem:Transversal}Let  \Enleve{$F=\{e_v | v \in V(G)\}$}\Ajoute{$F$}
be a set of edges of $G$, where each vertex of $V(G)$ appears
exactly once as the end of an edge of $F$. Then there exists a
normal partition $\mathcal T$ such that \Enleve{$F=\{e_T(v) |v \in V(G)\}$} \Ajoute{$F$ is the set of marked edges associated to $\mathcal T$}
if and only if $F$ is a transversal of the cycles of $G$.
\end{Thm}

\begin{Prf}
Let $\mathcal T$  be a normal partition, the set of marked edges
$\{e_T(v) |v \in V(G)\}$ is obviously a transversal of the cycles of
$G$, since $\mathcal T$ is partitioned into trails. Conversely,
assume that $F=\{e_v | v \in V(G)\}$ is a transversal of the cycles
of $G$.

Then the spanning subgraph  $G-F$  is a set of paths $\{P_1, P_2,
\ldots ,P_k\}$ (some of them being eventually reduced to a vertex).
Let $u_i$ and $v_i$ be the end vertices of $P_i$ ($1 \leq i \leq k$)
(when $P_i$ is reduced to a single vertex, we have $u_i=v_i$). We
add to each path $P_i$ the edges of $F$ which are incident to $u_i$
and $v_i$ and distinct from $e_{u_i}$ and $e_{v_i}$. We get
\Enleve{thus} a set of trails $\mathcal T=\{T_1,T_2, \ldots ,T_k\}$
which partition the edge set. We claim that $\mathcal T$ is a normal
partition. Indeed, let $v$ be any vertex of $G$. \Ajoute{The vertex}
$v$ is contained in some path $P_i$ of $G-F$ and $T_i$ must contain
the two edges incident to $v$ and distinct from the unique edge
associated to $v$ in $F$. Hence $v$ must be an internal vertex of
$T_i$ which implies that $v$ is normal.
\end{Prf}

\section{On compatible normal partitions}
\begin{Def}\label{Definition:CompatiblePartition}
Two partitions \stp  and \stprime of $E(G)$ into trails are {\em
compatible} when $e_T(v) \not= e_{T'}(v)$ for every vertex $v \in
V(G)$.
\end{Def}

\begin{Thm} \label{Theorem:PerfectmatchingOddNormalPartition}Let $G$ be a cubic graph. Then
the three following statements are equivalent.
\begin{itemize}
  \item[i)] $G$ has a perfect matching
  \item[ii)]$G$ has an odd normal partition
  \item[iii)] $G$ has two compatible normal partitions of length $3$
\end{itemize}
\end{Thm}

\begin{Prf}

Let $M$ be a perfect matching in $G$. Then $G-M$ is a $2-$factor of
$G$. Let us give any orientation to the cycles of this $2-$factor
and for each vertex $v$ let us denote the outgoing edge $o(v)$. For
each edge $e=uv \in M$, let $P_{uv}$ be the \Ajoute{trail} of length
$3$ obtained \Enleve{in concatenating} \Ajoute{by concatenation of
$o(u)$, $uv$ and $o(v)$.} Then $\mathcal T=\{P_{uv} | uv \in M \}$
is a normal odd partition (of length $3$) of $G$. \Ajoute{We obtain
a second normal partition $\mathcal T^{'}$ of length $3$, compatible
with $\mathcal T$, when we choose the other orientation on each
cycle.} Hence {\it(i)} implies {\it(ii)} and {\it(iii)}.

Let \np be a normal odd partition of $G$. For each trail $T_i \in T$
let us say that \Ajoute{an} edge $e$ of $T_i$ is {\em odd} whenever
the subtrails of $T_i$ obtained \Enleve{in} \Ajoute{by} deleting $e$
have odd lengths (an {\em even} edge being defined in the
\Enleve{obvious}\Ajoute{similar} way). \Enleve{A} \Ajoute{Any}
vertex $v \in V(G)$ is internal in exactly one trail of $\mathcal
T$. The edges of this trail being alternatively odd and even, $v$ is
incident to exactly one odd edge. Hence the odd edges so defined
induce a perfect matching of $G$ and {\it(ii)} implies {\it(i)}.

Since  {\it(iii)} implies obviously {\it(ii)}, the proof is
complete.
\end{Prf}

\Enleve{Compatible perfect path double covers When we can find two
compatible normal path partitions in a cubic graph we have, in fact
a particular $PPDC$ of its edge set.}

\begin{Def}\label{Definition:PPDC}
A {\em Perfect Path Double Cover} ($PPDC$ for short) is a collection
$\mathcal P$ of paths such that each edge of $G$ belongs to exactly
two members of $\mathcal P$ and each vertex occurs exactly twice as
an end path of $\mathcal P$.
\end{Def}

This notion has been introduced by Bondy (see \cite{Bon1}) who
conjectured that every simple graph admits a $PPDC$. This conjecture
was proved by Li \cite{Li}. When dealing with two compatible normal
\textbf{path} partitions $\mathcal P$ and $\mathcal P'$  in a cubic
graph, we have a particular $PPDC$. Indeed  every edge belongs to
exactly one path of $\mathcal P$ and one path of $\mathcal P'$ and
every vertex occurs exactly once as an end vertex of a path in
$\mathcal P$ and \Enleve{as well as an end vertex of} a path in
$\mathcal P'$. The qualifying adjective {\em compatible} says that
the two end edges are distinct for each vertex.

As a refinement of the notion of $PPDC$ we can define a $CPPDC$ for
a simple graph:
\begin{Def}\label{Definition:CPPDC}
A {\em Compatible Perfect Path Double Cover} ($CPPDC$ for short) is
a collection $\mathcal P$ of paths such that each edge of $G$
belongs to exactly two members of $\mathcal P$ and each vertex
occurs exactly twice as an end path of $\mathcal P$ and these two
ends are distinct.
\end{Def}

A natural question is thus to know which graphs admits a $CPPDC$. If
we restrict ourself to connected graphs, we immediately can see that
as soon as a graph \Ajoute{has} a \Ajoute{pendent} edge, a $CPPDC$
does not exist. We need thus to consider graphs with a certain
connectivity condition. \Ajoute{As an easy result we see that a
minimal $2-$edge connected graph has $CPPDC$.}

\begin{Prop}\Ajoute{Let $G$ ba a minimal $2-$edge connected simple graph. Then $G$
admits a $CPPDC$.}
\end{Prop}
\begin{Prf}
By induction on the number of vertices. The assertion can be
verified on the complete graph with three vertices, so assume that
$G$ has at least four vertices. It is well known (see Halin
\cite{Hal}) that $G$ contains a vertex $v$ whose degree is $2$. Let
$v_{1}$ and $v_{2}$ be the two neighbors of $v$.

\textbf{ case 1: $v_{1}v_{2} \in E(G)$.}

Let $G'$ be the graph obtained from $G$ by deleting $v$ and the edge
$v_{1}v_{2}$. Since $G$ is minimal $2-$edge connected, $G'$ has $2$
connected component $C_{i}$ ($i=1,2$), with $v_{i} \in C_{i}$. We
can see that these subgraphs are minimal $2-$edge connected. We can
thus find a $CPPDC$ $\mathcal T_{i}$  ($i=1,2$) for each of them.
Let $Q_{i},R_{i} \in \mathcal T_{i}$ ($i=1,2$) be the two paths with
end vertices $v_{i}$. Let $T_{1}=Q_{1}+v_{1}v_{2}v$ and
$T_{2}=Q_{2}+v_{2}v_{1}v$. Then $\mathcal T = \mathcal T_{1}
-Q_{1}+\mathcal T_{2} - Q_{2}+\{T_{1},T_{2}\}+v_1vv_2$ is a $CPPDC$ of $G$.

\textbf {case 2: $v_{1}v_{2} \not \in E(G)$ and $G-v$ is not minimal
$2-$edge connected.}

Let $G'$ be the graph obtained from $G$ by adding the edge
$v_{1}v_{2}$ and deleting the vertex $v$.

Assume that  $G'$ is still a   minimal $2-$edge connected. Then let
$\mathcal T'$ be a $CPPDC$ of $G'$ and let $T'_{1},T'_{2} \in
\mathcal T'$ be the two paths using the edge $v_{1}v_{2}$. We can
transform this $CPPDC$ of $G'$ in a $CPPDC$ of $G$ when we consider
$\mathcal T=\mathcal T' -\{T'_{1},T'_{2}\}+
\{T^{1}_{1},T^{2}_{1},T_{2}\}$ where $T_{2}$ is obtained from
$T'_{2}$ by inserting $v$ between $v_{1}$ and $v_{2}$ and
$T^{1}_{1}$ , $T^{2}_{1}$ are obtained from $T'_{1}$ by deleting the
edge $v_{1}v_{2}$ and adding the edge $v_{1}v$ to the subpath of
$T'_{1}$ containing $v_{1}$ (respectively, the edge $v_{2}v$ to the
subpath of $T'_{2}$ containing $v_{2}$).

When $G'$ is not a   minimal $2-$edge connected graph, there is an
edge of $G'$ whose deletion preserves the $2-$edge connectivity. In
fact, we can check that the only edge with that property must be the
edge $v_{1}v_{2}$ (otherwise $G$ itself is not minimal $2-$edge
connected). A contradiction since we have supposed that $G-v$ is not
minimal $2-$edge connected.

\textbf{ case 3: $v_{1}v_{2} \not \in E(G)$ and $G-v$ is minimal
$2-$edge connected.}

Let $G'=G-v$ and let $\mathcal T'$ be a $CPPDC$ of $G'$. Let
$Q_{i},R_{i} \in \mathcal T'$ ($i=1,2$) be the two paths with end
vertices $v_{i}$. We can consider that $Q_{1}$ and $Q_{2}$ are two
distinct paths of $\mathcal T'$. Then, let $\mathcal T=\mathcal
T'-\{Q_{1},Q_{2}\}+\{T_{1},T_{2}\}+v_{1}vv_{2}$ where $T_{i}$ is
obtained by concatenation of $Q_{i}$ and $v_{i}v$ ($i=1,2$). We can
check that $\mathcal T$ is a $CPPDC$ of $G$.

\end{Prf}

\Enleve{And} We propose as an open Problem

\begin{Problem}\label{Conjecture:CPPDC}
Every $2$-edge connected simple graph admits a $CPPDC$.
\end{Problem}

\begin{Rem}\label{Remark:CPPDC}Assume that a connected graph $G$ admits $CPPDC$. In doubling every
edge $e$ in $e'$ and $e''$ (let $G_2$ the graph so obtained), this
$CPPDC$ leads to an euler tour of $G_2$. This euler tour is
compatible (in the sense given by Kotzig \cite{Kot}) with the set of
transitions defined by $e'$ and $e''$ in each vertex.
\end{Rem}

\section{On three compatible  normal partitions}
We shall say that $G$ has three compatible normal partitions
\threenc whenever these partitions are pairwise compatible.

\Enleve{Remplacer $"$ par $''$ partout}

\Enleve{Remplacer $3$ compatible par three compatible}

\textbf{NB:} As usual $N(v)$ denotes the set of vertices adjacent to $v$. 
\begin{Thm} \label{Theorem:ThreeCompatibleGeneral} A
cubic graph  $G$ has three compatible normal partitions if and only
if $G$ has no loop.
\end{Thm}

\begin{Prf}
Let $G$ be a cubic graph with three compatible normal partitions
\threenc. Assume that $G$ contains a loop $vv$, let $w\neq v$ be the vertex adjacent to $v$.
Then one of these normal partitions, say $\mathcal T$, would be such that
$e_{\mathcal T}(v)=vw$. In that case $vv$ would be the trail
containing $v$ as an internal vertex, impossible.

Conversely, assume that $G$ has no loop and $G$ can not be provided
with three compatible normal partitions. We can suppose that $G$ has
been chosen with the minimum number of vertices for that property.
Figure \ref{Figure:ThetaGraph} shows that $G$ has certainly at least
$4$ vertices.

\begin{figure}[htb]
\centering \epsfsize=0.4 \hsize \noindent \epsfbox{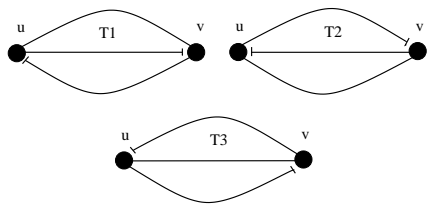}
\caption{Cubic graph on $2$ vertices with three compatible normal
partitions} \label{Figure:ThetaGraph}
\end{figure}

\begin{Clm} If $u$ and $v$ are joined by two edges $e_1$ and $e_2$, then there is a
third vertex $w$ adjacent to $u$ and $v$.
\end{Clm}
\begin{PrfClaim} Assume that $u$ is adjacent to $u'$ and $v$ to $v'$
with $u' \neq u$ and $v' \neq v$. Let $G'$ be the cubic graph
obtained from $G$ \Enleve{in} \Ajoute{by} deleting $u$ and $v$ and
joining $u'$ and $v'$ by a new edge. $G'$ is obviously a cubic graph
with no loop and $|V(G)| < |V(G')|$. We can thus find three
compatible normal partitions \threenc in $G'$.

The edge $u'v'$ of $G'$ is contained into $T \in \mathcal T$, $T'
\in \mathcal T'$ and $T^{''} \in \mathcal T^{''}$. For convenience,
$T_{1}$ and $T_{2}$ will \Ajoute{be} the subtrails of $T$ we have
obtained \Enleve{in} \Ajoute{by} deleting $u'v'$, with $u'$ an end
of $T_1$ and $v'$ an end of $T_2$. Following the same trick we get
$T'_{1}$ and $T'_{2}$, $T^{''}_{1}$ and $T^{''}_{2}$ when
considering $T'$ and $T^{''}$. It can be noticed that some of these
subtrails may have length $0$, which means that, following the
cases, $uv$ is the marked edge associated to $u$ or (and) $v$ in
$\mathcal T$, $\mathcal T'$ or $\mathcal T^{''}$.

Let $P_1=T_1+u'u$, $P_2=T_2+v've_1ue_2v$ and $\mathcal Q=\mathcal T
- P + \{P_1,P_2\}$. We can easily check that $\mathcal Q$ is a
normal partition of $G$ where $e_{\mathcal Q}(x)=e_{\mathcal T}(x) \
\forall x \neq u,v$ and $e_{\mathcal Q}(u)=uu'$, $e_{\mathcal
Q}(v)=e_2$.

In the same way, let $P'_1=T'_1+u'ue_2ve_1u$, $P'_2=T'_2+v'v$ and
$\mathcal Q'=\mathcal T' - P' + \{P'_1,P'_2\}$. Then $e_{\mathcal
Q'}(x)=e_{\mathcal T'}(x) \ \forall x \neq u,v$ and $e_{\mathcal
Q'}(u)=e_1'$, $e_{\mathcal Q'}(v)=vv'$. Hence $\mathcal Q'$ is a
normal partition compatible with $\mathcal Q$.

Finally, let $P^{''}_1=T^{''}_1+u'ue_1v$, $P^{''}_2=T'_2+v've_2u$
and $\mathcal Q^{''}=\mathcal T^{''} - P^{''} +
\{P^{''}_1,P^{''}_2\}$. Then $e_{\mathcal Q^{''}}(x)=e_{\mathcal
T^{''}}(x) \ \forall x \neq u,v$ and $e_{\mathcal Q^{''}}(u)=e_2$,
$e_{\mathcal Q^{''}}(v)=e_1$. Hence $\mathcal Q$, $\mathcal Q'$ and
$\mathcal Q^{''}$ are three compatible normal partitions of $G$, a
contradiction.
\end{PrfClaim}

\begin{Clm} if $uv \in E(G)$ then $|N(u)|=2$ or $|N(v)|=2$
\end{Clm}
\begin{PrfClaim}
Assume that $|N(u)|=3$ and $|N(v)|=3$ and let $u'$ and $u^{''}$ the
two neighbors of $u$ and $v'$ and $v^{''}$ those of $v$. Let $G'$ be
the graph obtained from $G$  by deleting $u$ and $v$ and joining
$u'$ and $u^{''}$ by a new edge as well as joining $v'$ and
$v^{''}$. $G'$ is obviously a cubic graph with no \Enleve{boucle}
\Ajoute{loop} and $|V(G)| < |V(G')|$. We can thus find three
compatible normal partitions \threenc in $G'$.

The edge $u'u^{''}$ of  $G'$ is contained into $T \in \mathcal T$,
$T' \in \mathcal T'$ and $T^{''} \in \mathcal T^{''}$ and we denote,
as in the previous claim by $T_1,T_2,T'_1,T'_2,T^{''}_1$ and
$T^{''}_2$ the subtrails of $T,T'$ and $T^{''}$ obtained \Enleve{in}
\Ajoute{by} deleting $u'u^{''}$ (with $u'$ an end of trails with
subscript $1$ and $u^{''}$ an end of trails with subscript $2$). If
$R \in \mathcal T$, $R' \in \mathcal T'$ and $R^{''} \in \mathcal
T^{''}$ are the
 trails using $v'v^{''}$, we can define also $R_1,R_2,R'_1,R'_2,R^{''}_1$ and
 $R^{''}_2$.

 We are going to construct three normal partition $\mathcal Q$,
$\mathcal Q'$ and $\mathcal Q^{''}$ of $G$ \Enleve{in} \Ajoute{by}
transforming locally $\mathcal T$, $\mathcal T'$ and $\mathcal
T^{''}$ in such a way that $e_{\mathcal Q}(x)=e_{\mathcal T}(x)$
$e_{\mathcal Q'}(x)=e_{\mathcal T'}(x)$ and $e_{\mathcal
Q^{''}}(x)=e_{\mathcal T^{''}}(x)$ $\forall x \neq u,v$. The
verification of this point, left to the reader, is immediate.

Let $P^{''}_1=T^{''}_1+u'uu^{''}+T^{''}_2$,
$P^{''}_2=R^{''}_1+v'vv^{''}+R^{''}_2$ and $P^{''}_3=uv$. $\mathcal
Q^{''}$ is then $\mathcal T^{''} - \{P^{''},R^{''}\} +
\{P^{''}_1,P^{''}_2,P^{''}_3\}$. We can remark that we have
subdivided $P^{''}$ and $R^{''}$ an we have add\Ajoute{ed} a trail
of length one ($uv$). We have hence, $e_{\mathcal Q^{''}}(u)=uv$ and
$e_{\mathcal Q^{''}}(v)=uv$.

 It must
be clear that we may have $T=R$ in $\mathcal T$, which means that
$u'u^{''}$ and $v'v^{''}$ are contained in the same trail of
$\mathcal T$. But we certainly have either $T_1 \neq R_1$ or $T_1
\neq R_2$ since $R_1$ and $R_2$ are two disjoint trails. Let us
consider the following partitions of the edge set of $G$:

$$\mathcal Q_1=\mathcal T - \{T_1,T_2\}+\{T_1+u'uvv'+R_1,T_2+u^{''}u,R_2+v^{''}v\}$$
$$\mathcal Q_2=\mathcal T - \{T_1,T_2\}+\{T_1+u'uvv^{''}+R_2,T_2+u^{''}u,R_2+v'v\}$$
$$\mathcal Q_3=\mathcal T - \{T_1,T_2\}+\{T_1+u'u,R_1+v'vuu^{''}+T_2,R_2+v^{''}v\}$$
$$\mathcal Q_4=\mathcal T - \{T_1,T_2\}+\{T_1+u'u,R_1+v'v,T_2+u^{''}uvv^{''}+R_2\}$$

$\mathcal Q_1$ is a normal partition of $G$ as soon as $T_1 \neq
R_1$ and  we can check, in that case, that $\mathcal Q_2$, $\mathcal
Q_3$ and $\mathcal Q_4$ are normal partitions of $G$. In the same
way, $\mathcal Q_2$ is a normal partition of $G$ as soon as $T_1
\neq R_2$ and  we can check, in that case, that $\mathcal Q_1$,
$\mathcal Q_3$ and $\mathcal Q_4$ are normal partitions of $G$.
$\mathcal Q_3$ is a normal partition of $G$ as soon as $T_2 \neq
R_1$ and, in that case,  $\mathcal Q_1$, $\mathcal Q_2$ and
$\mathcal Q_4$ are normal partitions of $G$. $\mathcal Q_4$ is a
normal partition of $G$ as soon as $T_2 \neq R_2$ and, in that case,
$\mathcal Q_1$, $\mathcal Q_2$ and $\mathcal Q_3$ are normal
partitions of $G$.

We can define analogously $\mathcal Q'_1$, $\mathcal Q'_2$,
$\mathcal Q'_3$ and $\mathcal Q'_4$ when considering $\mathcal T'$.

We can check moreover that these normal partitions (when they are
well defined) $\mathcal Q_1$, $\mathcal Q_2$, $\mathcal Q_3$,
$\mathcal Q_4$, $\mathcal Q'_1$, $\mathcal Q'_2$, $\mathcal Q'_3$
and $\mathcal Q'_4$ are compatible with $\mathcal Q^{''}$ since
$$e_{\mathcal Q_i}(u)=uu' \ {\rm or} \ e_{\mathcal Q_i}(u)=uu^{''} \ i=1,2,3,4$$
$$e_{\mathcal Q_i}(v)=vv' \ {\rm or} \ e_{\mathcal Q_i}(v)=vv^{''} \ i=1,2,3,4$$
$$e_{\mathcal Q'_i}(u)=uu' \ {\rm or} \ e_{\mathcal Q'_i}(u)=uu^{''} \ i=1,2,3,4$$
$$e_{\mathcal Q'_i}(v)=vv' \ {\rm or} \ e_{\mathcal Q'_i}(v)=vv^{''} \ i=1,2,3,4$$
We can verify that in each case to be considered with $\mathcal T$
($T_1=R_1$ and $T_2 \neq R_2$, $T_2=R_2$ and $T_1 \neq R_1$,
$T_1=R_2$ and $T_2 \neq R_1$, $T_2=R_1$ and $T_1 \neq R_2$, $T_1,
T_2, R_1, R_2$ all distinct) together with  the similar cases for
$\mathcal T'$ we can choose a normal partition $\mathcal Q$ in
$\{\mathcal Q_1, \mathcal Q_2, \mathcal Q_3, \mathcal Q_4\}$ and a
normal partition $\mathcal Q'$ in $\{\mathcal Q'_1, \mathcal Q'_2,
\mathcal Q'_3, \mathcal Q'_4\}$ which are compatible and hence three
normal partitions compatible $\mathcal Q, \mathcal Q'$ and $\mathcal
Q^{''}$ for $G$, a contradiction.
\end{PrfClaim}


Assume that $u$ and $v$ are joined by two edges in $G$, then, from
Claim 1, there is unique new vertex $w$ joined to $u$ and $v$. This
vertex is adjacent to $x \neq u,v$ which have itself a neighbor $z
\neq u,v$. Since $|N(w)|=2$, by Claim 2, $N(x)=\{w,z\}$. The
vertices $x$ and $z$ being joined by two edges, $x$ and $z$ must
have a common neighbor by Claim 1, impossible. Hence $G$ does not
exist and the proof is complete.
\end{Prf}

\begin{Prop} \label{Proposition:EdgeStructure}Let $G$ be a cubic graph
having three compatible normal partitions then every edge  $e \in
E(G)$ verifies exactly one of the followings

\begin{itemize}
  \item  $e$ is an internal edge in exactly one partition
  \item $e$ is an internal edge in exactly two partitions
\end{itemize}
Moreover, in the second case, the edge $e$ itself is a trail of the
third partition.
\end{Prop}

\begin{Prf} Let $e=xy$ be any edge of $G$ and let $\mathcal T$, $\mathcal T'$ and $\mathcal T^{''}$
 \Ajoute{be} three compatible normal partitions.
 If $e$ is not an internal edge in $\mathcal T$, $\mathcal T'$ nor $\mathcal T^{''}$ then $e$ is an end
 edge for a trail of
$\mathcal T$, $\mathcal T'$ and $\mathcal T^{''}$. In $x$ or $y$ we
should have two partitions (say $\mathcal T$ and $\mathcal T'$) for
which $e_{\mathcal T}(x)=e_{\mathcal T'}(x)$ ($e_{\mathcal
T}(y)=e_{\mathcal T'}(y)$ respectively), a contradiction. \Ajoute{So
let us suppose} that $e$ is an internal edge in $\mathcal T$,
$\mathcal T'$ and $\mathcal T^{''}$. Let $a$ and $b$ the two other
neighbors of $x$. We should have then
\begin{itemize}
  \item $e_{\mathcal T}(x)=xa$ or $xb$
  \item $e_{\mathcal T'}(x)=xa$ or $xb$
  \item $e_{\mathcal T^{''}}(x)=xa$ or $xb$
\end{itemize}
which is impossible since the three partitions are compatible.
Assume now that $e$ is an internal edge of a trail in $\mathcal T$
and in $\mathcal T'$ and let $a$ and $b$ the two other neighbors of
$x$. Up to the names of vertices we have
\begin{itemize}
  \item $e_{\mathcal T}(x)=xa$
  \item $e_{\mathcal T'}(x)=xb$
\end{itemize}
From the third partition $\mathcal T^{''}$, we must have
$e_{\mathcal T^{''}}(x)=xy$. In the same way we should obtain
$e_{\mathcal T^{''}}(y)=yx$. Hence the trail containing $e=xy$ is
reduced to $e$, as claimed.\end{Prf}

It can be noticed that whenever a cubic graph can be provided with
three compatible normal partitions at least one edge is the internal
edge in exactly one partition.

\begin{Prop} \label{Proposition:AtLeastOnedegIsolated}Let $G$ be a cubic graph
having three compatible normal partitions. Then at least one edge  $e
\in E(G)$ is the internal edge in exactly one partition.

\end{Prop}
\begin{Prf}
Let $\mathcal T$, $\mathcal T'$ and $\mathcal T^{''}$ be
 three compatible normal partitions of $G$. The set of trails of length $1$ in $\mathcal T$ is a
matching of $G$ which means that $\mathcal T$ has at most
$\frac{n}{2}$ such trails. If each edge of $G$ is the internal edge
in exactly two partitions we must have
$$ |E(G)|=n^1_{\mathcal T} + n^1_{\mathcal T'}+ n^1_{\mathcal T^{''}} \leq
3\frac{n}{2}= |E(G)|$$ Hence the set of edges which are trails of
length $1$ in $\mathcal T$ is a perfect matching $M$ of $G$. In that
case, the set of marked edges associated to $\mathcal T$ is
precisely this set $M$, which is not transversal of the cycles of
$G$, a contradiction with Theorem \ref{Theorem:Transversal}.
\end{Prf}

\begin{Thm} \label{Theorem:ThreeCompatibleTrailsindex3}Let $G$ be a
\cubsimplethree then $G$ has three compatible normal partitions
\threenc such that
\begin{itemize}
  \item $\mathcal T$ is odd
  \item $\mathcal T'$ has length $3$
  \item $\mathcal T^{''}$ has length $4$
\end{itemize}

\end{Thm}

\begin{Prf}
\Enleve{Avons nous besoin de faire la distinction? Oui puisque le
résultat sur les couplages forts est vrai pour les graphes simples
}\Enleve{We shall prove first this result for simple graphs}In
\cite{FouThuVanWoj}, it is proved that, given a $3$-edge colouring
of $G$ with $\alpha$,$\beta$ and $\gamma$ then there exists a strong
matching intersecting every cycle belonging to the 2-factor induced
by the two colours $(\alpha$ and $\beta)$. Assume that $\mathcal
C=\{C_1,C_2, \ldots C_k\}$ is such \Ajoute{a} $2-$factor
($G-\mathcal C$ is a perfect matching) and let $F=\{u_iv_i \in C_i|
\ 1\leq i \leq k$) (minimal for the inclusion) be a strong matching
intersecting  \Ajoute{each cycle of} this $2-$factor.

For each $u_iv_i \in F$, $x_i$ is the vertex in the neighborhood of
$u_i$ which is not one of its  neighbor (\Ajoute{predecessor or
successor}) on $C_i$ while $y_i$ is defined similarly for $v_i$
(note that $x_i$ and $y_i$ may be vertices of $C_i$ or not). Let
$T_i$ be the trail obtained from $C_i$ \Enleve{in} \Ajoute{by}
adding the edge $u_ix_i$ and considering that this trail ends with
$v_iu_i$ (Note that $u_i$ is an internal vertex of $T_i$).

Let $\mathcal T$ be the trail partition containing every trail $T_i$
($1\leq i \leq k$) and all the edges of the perfect matching
$G-\mathcal C$ which are not in some $T_i$. We can check that
$\mathcal T$ is a normal odd partition for which the
following\Enleve{s} \Ajoute{holds}

\begin{itemize}
  \item $e_{\mathcal T}(u_i)=u_iv_i$
  \item $e_{\mathcal T}(x_i)=x_iu_i$
  \item $e_{\mathcal T}(v)$ is the edge of $G-\mathcal C$ \Ajoute{for each
  vertex $v \not=u_i,v_i$}
\end{itemize}

We construct now \Ajoute{the trail partition} $\mathcal T'$.
\Ajoute{Let us give the orientation} to each cycle of $\mathcal C$.
This orientation is such that the successor of $u_i$ is  $v_i$. For
each vertex $v$, $o(v)$ denotes the successor of $v$ in that
orientation and $p(v)$ its predecessor. As in \Ajoute{the proof of}
Theorem \ref{Theorem:PerfectmatchingOddNormalPartition} we get hence
a normal partition $\mathcal T'$ where each trail is a path of
length $3$. Moreover $e_{\mathcal T'}(v)=vp(v)$ \Ajoute{for every
vertex $v$}.

Before constructing $\mathcal T^{''}$, we construct $\mathcal
T^{'''}$ \Ajoute{by} using the reverse orientation on each cycle of
$\mathcal C$. This normal partition of length $3$ is such that
$e_{\mathcal T'''}(v)=vo(v)$.

For each vertex $v \not = u_i \ 1\leq i \leq k$ we have $e_{\mathcal
T}(v) \not = e_{\mathcal T'}(v) \not = e_{\mathcal T'''}(v)$.

For $v  = u_i \ 1\leq i \leq k$, we have $e_{\mathcal
T}(u_i)=u_iv_i$, $e_{\mathcal T'}(u_i)=u_ip(u_i))$ (where $p(u_i)
\not = v_i$) and $e_{\mathcal T'''}(u_i)=u_iv_i$. Since $e_{\mathcal
T}(u_i)=e_{\mathcal T^{'''}}(u_i)$, $\mathcal T$ and $\mathcal
T^{'''}$ are not compatible.

Our goal now is to proceed  to switchings on $\mathcal T^{'''}$ in
each vertex $u_i$ in order to get $\mathcal T^{''}$ where these
incompatibilities are dropped. For this purpose, \Ajoute{we extend
every path of} length $3$ of $\mathcal T^{'''}$ ending with $v_iu_i$
\Enleve{is augmented} with the edge $u_ip(u_i)$. We get hence of
path of length $4$ and, since $F$ is a strong matching, we are sure
that we cannot extend this path in the other direction. The path of
$\mathcal T^{'''}$ ending with $u_ip(u_i)$ is shorten \Enleve{in} \Ajoute{by} deleting
the edge $u_ip(u_i)$, we get hence of path of length $2$ ending with
$x_iu_i$, and we are sure that this path cannot be shorten at the
other end, since $F$ is a strong matching. Let $\mathcal T^{''}$ be
the partition so obtained. $\mathcal T^{'''}$ being normal and
$\mathcal T^{''}$ having the same number of trails $\mathcal T^{''}$
is also normal by Proposition \ref{Proposition:NormalNover2}.

For each vertex $v \not = u_i \ 1\leq i \leq k$, $e_{\mathcal
T'''}(v)=e_{\mathcal T^{''}}(v)$ and we have thus $e_{\mathcal T}(v)
\not = e_{\mathcal T'}(v) \not = e_{\mathcal T^{''}}(v)$. For $v  =
u_i \ 1\leq i \leq k$, we have $e_{\mathcal T}(u_i)=u_iv_i$,
$e_{\mathcal T'}(u_i)=u_ip(u_i)$  and $e_{\mathcal
T^{''}}(u_i)=u_ix_i.$

\threenc are thus compatible ,$\mathcal T$ is odd, $\mathcal T'$ has
length $3$ and $\mathcal T^{''}$ has length $4$ as claimed.

\Enleve{Bon, et maintenant le cas des graphes ayant des arêtes
multiples?}
\end{Prf}

In fact we can extend the result to cubic graphs with multiple
edges.
\begin{Thm} \label{Theorem:ThreeCompatibleTrailsMultipleIndex3}Let $G$ be a
\cubthree then $G$ has three compatible normal partitions \threenc
such that
\begin{itemize}
  \item $\mathcal T$ is odd
  \item $\mathcal T'$ has length $3$
  \item $\mathcal T^{''}$ has length at most $4$
\end{itemize}

\end{Thm}

\begin{Prf}
By induction on the number of vertices of $G$. In Figure
\ref{Figure:ThetaGraph} we can see that the result holds for the
cubic graph with two vertices and three edges. If $G$ is simple, we
are done by Theorem \ref{Theorem:ThreeCompatibleTrailsindex3}. So
assume that $G$ has at least $4$ vertices and let $u$ and $v$ be two
vertices joined by \Enleve{tow}\Ajoute{two} edges $e_{1}$ and $e_{2}$. Let $x$ be the
third vertex adjacent to $u$ and $y$ the one adjacent to $v$. Let
$G'$ be the graph obtained from $G$ by deleting $u$ and $v$ and
adding a new edge $e$ between $x$ and $y$. From the hypothesis of
induction, let $\mathcal Q$, $\mathcal Q'$ and $\mathcal Q^{''}$  be
three compatible normal partitions of $G'$. We have to discuss three
cases following the fact that $e$ is in $\mathcal Q$, $\mathcal Q'$
or $\mathcal Q^{''}$

\textbf{case 1: $e$ is an internal edge of a trail $Q \in \mathcal
Q$}

In that case $e$ is an end edge of a trail $Q' \in \mathcal Q'$ as
well as an end edge of a trail $Q^{''} \in \mathcal Q^{''}$. Without
loss of generality, we assume that $e_{\mathcal Q'}(x)=xy$ and
$e_{\mathcal Q^{''}}(y)=yx$. Hence $Q'$ and $Q^{''}$ end both with
the edge $xy$. Let $T$ be the trail obtained from $Q$ by deleting
the edge $xy$ and adding the path $xue_{1}vy$ (the notation
$ue_{1}v$ means that we use explicitly the edge $e_{1}$ in order to
connect $u$ and $v$). Let $T'$ be the trail obtained from $Q'$ by
deleting the edge $xy$ and adding the edge $yv$. Let $T^{''}$ be the
trail obtained from $Q^{''}$ by deleting the edge $yx$ and adding
the edge $xu$. Then we can construct \threenc three compatible
normal partitions of $G$ in the following way:
\begin{itemize}
  \item \Ajoute{$\mathcal T=\mathcal Q -Q + T+ue_2v$}
  \item $\mathcal T'=\mathcal Q' -Q' + T' +xue_{2}ve_{1}u$
  \item $\mathcal T^{''}=\mathcal Q^{''}-Q^{''}+T^{''}+yve_{2}ue_{1}v$
\end{itemize}
We can check that the conditions on the lengths are verified for
\threencsv.

\textbf{case 2: $e$ is an internal edge of a trail $Q' \in \mathcal
Q'$}

 In that case $e$ is an end edge of a trail $Q \in \mathcal Q$
as well as an end edge of a trail $Q^{''} \in \mathcal Q^{''}$.
Without loss of generality, we assume that \Enleve{$e_{\mathcal Q}(x)=xy$}\Ajoute{$e_{\mathcal Q}(y)=xy$}
and \Enleve{$e_{\mathcal Q^{''}}(y)=yx$}\Ajoute{$e_{\mathcal Q^{''}}(x)=yx$}. Hence $Q$ and $Q^{''}$ end both
with the edge $xy$. Let us recall that $Q'$ has length $3$. Let $zx$
and $ty$ be the end edges of $Q$. Let $T^{''}$ be the trail obtained
from $Q^{''}$ by deleting the edge $yx$ and adding the edge $yv$.
Let $T$ be the trail obtained from $Q$ by deleting the edge $xy$ and
adding the path $xue_{1}vy$

Then we can construct \threenc three compatible normal partitions of
$G$ in the following way:
\begin{itemize}
  \item $\mathcal T=\mathcal Q -Q + T+xue_{1}ve_{2}u$
  \item $\mathcal T'=\mathcal Q' -Q'  +zxue_{2}v+tyve_{1}u$
  \item $\mathcal T^{''}=\mathcal Q^{''}-Q^{''}+T^{''}+yve_{2}ue_{1}v$
\end{itemize}
We can check that the conditions on the lengths are verified for
\threencsv.

\textbf{case 3: $e$ is an internal edge of a trail $Q^{''} \in
\mathcal Q^{''}$}

A similar technique can be used to solve this case.

\end{Prf}

\begin{Thm} \label{Theorem:Bipartite}Let $G$ be a cubic graph. Then
the following\Enleve{s} \Ajoute{statements} are equivalent
\begin{itemize}
  \item [i)]$G$ can be provided with three compatible normal  partitions of length $3$
  \item [ii)]$G$ can be provided with three compatible normal  odd partitions
  where each edge is an internal edge in exactly one partition
  \item [iii)] $G$ is bipartite
\end{itemize}
\end{Thm}

\begin{Prf} Assume first that $G$ can be provided with  three compatible normal partitions of
length $3$, say $\mathcal T$, $\mathcal T'$ and $\mathcal T^{''}$.
Since the \Ajoute{average} \Enleve{mean} length of each partition is
$3$ (Proposition \ref{Proposition:Distribution}), each trail of each
partition has length \Enleve{exactly} $3$. $\mathcal T$, $\mathcal
T'$ and $\mathcal T^{''}$ are thus three normal odd partitions and
from Proposition \ref{Proposition:EdgeStructure}, each edge is the
internal edge of one trail in exactly one partition. Conversely
assume that $G$ can be provided with three compatible normal odd
partitions  where each edge is an internal edge in exactly one
partition. Then, \Ajoute{by} Proposition
\ref{Proposition:EdgeStructure} there is no trail of length $1$ in
any of these partitions. Since the \Enleve{mean} \Ajoute{average}
length of each partition is $3$, that means that each trail in each
partition has length  \Enleve{exactly} $3$. Hence $(i) \equiv (ii)$.

We prove now that $(i) \equiv (iii)$. Let $\mathcal T$, $\mathcal
T'$ and $\mathcal T^{''}$ three compatible normal partitions of
length $3$. Following the proof of Theorem
\ref{Theorem:PerfectmatchingOddNormalPartition} the internal edges
of trails of $\mathcal T$ ($\mathcal T'$ and $\mathcal T^{''}$
respectively) constitute a perfect matching (say $M$ $M'$ and
$M^{''}$ respectively).

Let $a_0a_1a_2a_3$ be a trail of $\mathcal T$ and let $b_1$ and
$b_2$ the third neighbors of $a_1$ and $a_2$ respectively. By
definition, we have $e_{\mathcal T}(a_1)=a_1b_1$ and $e_{\mathcal
T}(a_2)=a_2b_2$.

Since $a_0a_1$ and $a_2a_3$ must be internal edges in a trail of
$\mathcal T'$ or (exclusively)  $\mathcal T^{''}$, \Ajoute{without
loss of generality we may assume} that $a_0a_1$ is an internal edge
of a trail $T_1'$ of $\mathcal T'$. $T_1'$ does not use $a_1a_2$
otherwise $e_{\mathcal T'}(a_1)=a_1b_1$, a contradiction with
$e_{\mathcal T}(a_1)=a_1b_1$ since $\mathcal T$ and $\mathcal T'$
are compatible. Hence $T'_1$ uses $a_1b_1$ and $e_{\mathcal
T'}(a_1)=a_1a_2$.

Assume now that $a_2a_3$ is an internal edge of a trail $T'_2$ of
$\mathcal T'$. Reasoning in the same way, we get that $e_{\mathcal
T'}(a_2)=a_2a_1$.
 These two results leads to the fact that $a_1a_2$ must be a trail
 in $\mathcal T'$, which is impossible since each trail has length
 \Enleve{exactly} $3$.

 Hence, whenever $a_0a_1$ is supposed to be an internal edge in a
 trail of $\mathcal T'$, we must have $a_2a_3$ as an internal edge
 in a trail of $\mathcal T^{''}$. The two internal vertices of
 $a_0a_1a_2a_3$ can be thus distinguished, following the fact that
 the end edge of $\mathcal T$ to whom they are incident is internal in
 $\mathcal T'$ (say {\em red} vertices) or $\mathcal T^{''}$ (say {\em blue} vertices).
 The same holds for each trail in $\mathcal T$ (and incidently for each partition $\mathcal T'$ and
$\mathcal T^{''}$). The edge $a_1b_1$ as end-edge of $\mathcal T$
cannot be an internal edge in $T'$ since the trail of length $3$
going through $a_0a_1$ ends with $a_1b_1$. Hence $a_1b_1$ is an
internal edge in $\mathcal T^{''}$ and $b_1$ is a blue vertices.
Considering now $a_0$, this vertex is the internal vertex of a trail
of length $3$ of $\mathcal T$. Since $a_0a_1 \in M'$ and $M'$ is a
perfect matching, $a_0$ cannot be incident to an other internal edge
of a trail in $\mathcal T'$ and $a_0$ must be a blue vertex. Hence
$a_1$ is a red vertex and its neighbors are all blue vertices. Since
we can perform this reasoning in each vertex, $G$ is bipartite as
claimed.

\trou Conversely, assume that $G$ is bipartite and let
$V(G)=\{W,B\}$ be the bipartition of its vertex set. In the
following, a vertex in $W$ will be represented by a circle ($\circ$)
while a vertex in $B$ will be represented by a bullet ($\bullet$).
>From K\"{o}nig's theorem \cite{Kon} $G$ is a \cubthree. Let us
consider a coloring of its edge set with three colors $ \{\alpha,
\beta, \gamma\}$. Let us denote by $\alpha \bullet \beta \circ
\gamma$ a trail of length $3$ which is obtained in considering an
edge $uv$ ($u \in B$ and $v \in W$) colored with $\beta$ together
with the edge colored $\alpha$ incident with $u$ and the edge
colored with $\gamma$ incident with $v$. It can be easily checked
that the set $\mathcal T$ of $\alpha \bullet \beta \circ \gamma$
trails of length $3$ is a normal odd partition of length $3$. We can
define in the same way $\mathcal T'$ as the set of $\beta \bullet
\gamma \circ \alpha$ trails of length $3$ and $\mathcal T^{''}$ as
the set of $\gamma \bullet \alpha \circ \beta$ trails of length $3$.

Hence $\mathcal T$, $\mathcal T'$ and $\mathcal T^{''}$ is a set of
three normal odd partitions of length $3$. We claim that these
partitions are compatible. Indeed, let $v \in W$ be a vertex and
$u_1,u_2$ and $u_3$ its neighbors. Assume that $u_1v$ is colored
with $\alpha$, $u_2v$ is colored with $\beta$ and $u_3v$ is colored
with $\gamma$ . Hence $u_1v$ is internal in \Enleve{an} \Ajoute{a}
$\gamma \bullet \alpha \circ \beta$ trail of $\mathcal T^{''}$ and
$e_{\mathcal T^{''}}(v)=vu_3$. The edge $u_2v$ is internal in
\Enleve{an} \Ajoute{a} $\alpha \bullet \beta \circ \gamma$ trail of
$\mathcal T$ and $e_{\mathcal T}(v)=vu_1$. The edge $u_3v$ is
internal in \Enleve{an} \Ajoute{a} $\beta \bullet \gamma \circ
\alpha$ trail of $\mathcal T'$ and $e_{\mathcal T'}(v)=vu_2$. Since
the same reasoning can be performed in each vertex of $G$, the three
 partitions $\mathcal T$, $\mathcal T'$ and $\mathcal T^{''}$ are
compatible.

\end{Prf}

%
%

\begin{Thm} \label{Theorem:1Length3PlusTwoOdd}Let $G$ be a
  cubic graph with three compatible normal  partitions \threenc
such that
\begin{itemize}
  \item $\mathcal T$ has length $3$
  \item $\mathcal T'$ and $\mathcal T^{''}$ are odd
\end{itemize}
Then $G$ is a \cubthreev.
\end{Thm}
\begin{Prf}  Since $\mathcal T$ has length $3$, every trail of
$\mathcal T$  has length $3$. Hence there is no edge which can be an
internal edge of a trail of $\mathcal T'$ and a trail of $\mathcal
T^{''}$, since, by Proposition \ref{Proposition:EdgeStructure} such
an edge would be a trail of length $1$ in $\mathcal T$. The perfect
matchings associated to $\mathcal T'$ and $\mathcal T^{''}$ (see
Theorem \ref{Theorem:PerfectmatchingOddNormalPartition}) are thus
disjoint and induce an even $2$-factor of $G$, which means that $G$
is a \cubthreev, as claimed.

\end{Prf}

\begin{Prop} \label{Proposition:Triangle3oddCompatible} Let $G$
be a cubic graph which can be provided with three compatible normal
odd partitions. \Ajoute{Then the graph } $G'$  obtained \Ajoute{by}
replacing a vertex by a triangle, can also be provided with three
compatible normal odd partitions.
\end{Prop}
\begin{Prf}
Let $u$ be a vertex of $G$ and $v_1,v_2,v_3$ its neighbors (not
necessarily distinct). Assume that \threenc is a set of $3$
compatible normal odd partitions of $G$ such that, $e_{\mathcal
T}(u)=uv_1$, $e_{\mathcal T'}(u)=uv_2$ and $e_{\mathcal
T^{''}}(u)=uv_3$. Let $T_1$ and $T_2$ the two trails of $\mathcal T$
such that $u$ is an end of $T_1$ and an internal vertex of $T_2$.
$T^1_1$ ending in $v_1$, $T^2_1$ ending in $v_2$ and $T^2_2$ ending
in $v_3$ denote the subtrails of $T_1$ and $T_2$ obtained
\Enleve{in} \Ajoute{by} deleting $u$. We define similarly $T'^1_1$
ending in $v_2$, $T'^2_1$ ending in $v_1$ and $T'^2_2$ ending in
$v_3$ when considering $T'_1$ and $T'_2$ in $\mathcal T'$ as well as
$T^{''1}_1$ ending in $v_3$, $T^{''2}_1$ ending in $v_2$  and
$T^{''2}_2$ ending in $v_1$ when considering $T^{''}_1$ and
$T^{''}_2$ in $\mathcal T^{''}$.

When we transform $G$ in $G'$ the vertex $u$ is deleted and replaced
by the triangle $u_1,u_2,u_3$ with $u_i$ joined to $v_i$
($i=1,2,3$).

Let $\mathcal Q$, $\mathcal Q'$ and $\mathcal Q^{''}$ be defined in
$G'$ by

$$\mathcal Q= \mathcal T -\{T_1,T_2\}
+\{T^1_1+v_1u_1,T^2_1+v_2u_2u_1u_3v_3+T^2_2,u_2u_3\}$$
$$\mathcal Q'= \mathcal T' -\{T'_1,T'_2\}
+\{T'^1_1+v_2u_2,T'^2_1+v_1u_1u_2u_3v_3+T'^2_2,u_1u_3\}$$
$$\mathcal Q^{''}= \mathcal T^{''} -\{T^{''}_1,T^{''}_2\}
+\{T^{''1}_1+v_3u_3,T^{''2}_1+v_2u_2u_1u_3v_3+T^{''2}_2,u_2u_1\}$$

It is a routine matter to check that $\mathcal Q, \mathcal Q'$ and
$\mathcal Q^{''}$ are three compatible normal odd partitions.
\end{Prf}

It can be pointed out that cubic graphs with with three compatible
normal odd partitions are bridgeless.

\begin{Prop} \label{Proposition:Bridgeless3compatibleNormalOdd}Let $G$ be
a cubic graph with three compatible normal odd partitions. Then $G$
is bridgeless.
\end{Prop}
\begin{Prf}
Assume that $xy$ is a bridge of $G$ and let $C$ be the connected
component of $G-xy$ containing $x$. Since $G$ has three compatible
normal odd partitions, one of these partitions, say $\mathcal T$, is
such that $e_{\mathcal T}(x)=xy$. The edges of $C$ are thus
partitioned into odd trails (namely the trace of $\mathcal T$ on
$C$). We have $$m=|E(C)| = \frac{3(|C|-1)+2}{2}$$ and $m$ is even
whenever $|C| \equiv 3 \ mod \ 4$ while $m$ is odd whenever $|C|
\equiv 1 \ mod \ 4$. The trace of $\mathcal T$ on $C$ is a set of
$\frac{|C|-1}{2}$ trails and this number is odd when $|C| \equiv 3 \
mod \ 4$ and even otherwise. Hence, when $|C| \equiv 3 \ mod \ 4$ we
must have an odd number of odd trails partitioning $E(C)$ but, in
that case $m$ is even and when $|C| \equiv 1 \ mod \ 4$ we must have
an even number of odd trails partitioning $E(C)$ but, in that case
$m$ is odd, contradiction.
\end{Prf}

Fan and Raspaud \cite{FanRas} conjectured that any bridgeless cubic
graph can be provided with three perfect matching with empty
intersection.

\begin{Thm} \label{Theorem:3PerfectMatching}Let $G$ be a cubic graph with three compatible
normal odd partitions   then there exist $3$ perfect matching $M$,
$M'$ and $M^{''}$ such that $M \cap M' \cap M^{''} = \emptyset$.
\end{Thm}

\begin{Prf}Following the proof of Theorem
\ref{Theorem:PerfectmatchingOddNormalPartition} the odd edges of
trails of $\mathcal T$ ($\mathcal T'$ and $\mathcal T^{''}$
respectively) constitute a perfect matching (say $M$ $M'$ and
$M^{''}$ respectively). Let $v$ be any vertex and $u_1, u_2$ and
$u_3$ its neighbors. $\mathcal T$, $\mathcal T'$ and $\mathcal
T^{''}$ being compatible, we can suppose that $ e_{\mathcal
T}(v)=vu_1$, $e_{\mathcal T'}(v)=vu_2$ and $ e_{\mathcal
T^{''}}(v)=vu_3$. \Enleve{$vu_1$ is an end edge of a trail in
$\mathcal T$, this edge is not an odd edge in $\mathcal T$ and thus
$vu_1 \not \in M'$.} \Ajoute{Since $vu_1$ is an end edge of a trail
of $\mathcal T$, this edge is not an odd edge relatively to
$\mathcal T$. That means that $vu_{1} \not \in M$.} In the same way
$vu_2 \not \in M'$ and $vu_3 \not \in M^{''}$. Hence, any edge
incident to $v$ is contained in at most two perfect matchings among
$M, M'$ and $M^{''}$. Which means that $M \cap M' \cap M^{''} =
\emptyset$

\end{Prf}

Theorem \ref{Theorem:3PerfectMatching} above implies that
\Enleve{the} \Ajoute{Fan-Raspaud} Conjecture is true for graphs with $3$
compatible normal odd partitions. By the way, this conjecture seems
to be originated independently by Jackson. Goddyn \cite{God} indeed
mentioned this problem proposed by Jackson for $r-$graphs
($r-$regular graphs with an even number of vertices such that all
odd cuts have size at least $r$, as defined by Seymour \cite{Sey})
in the proceedings of a joint summer research conference on graphs
minors which dates back 1991.
\begin{figure}
\centering
\subfigure[]{\label{Figure:Petersen1}\setlength{\unitlength}{0.00021872in}
\begingroup\makeatletter\ifx\SetFigFont\undefined
\def\x#1#2#3#4#5#6#7\relax{\def\x{#1#2#3#4#5#6}}%
\expandafter\x\fmtname xxxxxx\relax \def\y{splain}%
\ifx\x\y   
\gdef\SetFigFont#1#2#3{%
  \ifnum #1<17\tiny\else \ifnum #1<20\small\else
  \ifnum #1<24\normalsize\else \ifnum #1<29\large\else
  \ifnum #1<34\Large\else \ifnum #1<41\LARGE\else
     \huge\fi\fi\fi\fi\fi\fi
  \csname #3\endcsname}%
\else
\gdef\SetFigFont#1#2#3{\begingroup
  \count@#1\relax \ifnum 25<\count@\count@25\fi
  \def\x{\endgroup\@setsize\SetFigFont{#2pt}}%
  \expandafter\x
    \csname \romannumeral\the\count@ pt\expandafter\endcsname
    \csname @\romannumeral\the\count@ pt\endcsname
  \csname #3\endcsname}%
\fi
\fi\endgroup
{\renewcommand{\dashlinestretch}{30}
\begin{picture}(6496,6365)(0,-10)
\thicklines
\path(4982,351)(4982,150)
\thinlines
\put(1488,175){\blacken\ellipse{336}{336}}
\put(1488,175){\ellipse{336}{336}}
\put(176,3823){\blacken\ellipse{336}{336}}
\put(176,3823){\ellipse{336}{336}}
\put(1968,3215){\blacken\ellipse{336}{336}}
\put(1968,3215){\ellipse{336}{336}}
\put(2528,1679){\blacken\ellipse{336}{336}}
\put(2528,1679){\ellipse{336}{336}}
\put(4160,1695){\blacken\ellipse{336}{336}}
\put(4160,1695){\ellipse{336}{336}}
\put(5296,255){\blacken\ellipse{336}{336}}
\put(5296,255){\ellipse{336}{336}}
\put(4720,3279){\blacken\ellipse{336}{336}}
\put(4720,3279){\ellipse{336}{336}}
\put(6320,3887){\blacken\ellipse{336}{336}}
\put(6320,3887){\ellipse{336}{336}}
\put(3232,6175){\blacken\ellipse{336}{336}}
\put(3232,6175){\ellipse{336}{336}}
\thicklines
\path(2194,3125)(2082,2963)
\path(1269,438)(157,3875)(1994,3200)
	(4644,3225)(6319,3925)(5257,263)
	(4157,1700)(2144,3063)
\path(4994,238)(1382,238)(2569,1675)
	(3294,4263)(3294,6200)(6144,4100)
\path(294,4050)(3007,6063)
\path(2969,6150)(3094,6025)
\path(244,4138)(357,4000)
\path(4544,3100)(2744,1838)
\path(3407,4088)(4132,1938)
\path(3357,4038)(3507,4100)
\path(4082,1913)(4219,1963)
\path(4507,3138)(4582,3050)
\path(2719,1913)(2819,1775)
\path(1357,500)(1182,388)
\path(6207,4163)(6082,4026)
\thinlines
\put(3280,4287){\blacken\ellipse{336}{336}}
\put(3280,4287){\ellipse{336}{336}}
\end{picture}
}}
\subfigure[]{\label{Figure:Petersen2}\setlength{\unitlength}{0.00021872in}
\begingroup\makeatletter\ifx\SetFigFont\undefined
\def\x#1#2#3#4#5#6#7\relax{\def\x{#1#2#3#4#5#6}}%
\expandafter\x\fmtname xxxxxx\relax \def\y{splain}%
\ifx\x\y   
\gdef\SetFigFont#1#2#3{%
  \ifnum #1<17\tiny\else \ifnum #1<20\small\else
  \ifnum #1<24\normalsize\else \ifnum #1<29\large\else
  \ifnum #1<34\Large\else \ifnum #1<41\LARGE\else
     \huge\fi\fi\fi\fi\fi\fi
  \csname #3\endcsname}%
\else
\gdef\SetFigFont#1#2#3{\begingroup
  \count@#1\relax \ifnum 25<\count@\count@25\fi
  \def\x{\endgroup\@setsize\SetFigFont{#2pt}}%
  \expandafter\x
    \csname \romannumeral\the\count@ pt\expandafter\endcsname
    \csname @\romannumeral\the\count@ pt\endcsname
  \csname #3\endcsname}%
\fi
\fi\endgroup
{\renewcommand{\dashlinestretch}{30}
\begin{picture}(6672,6301)(0,-10)
\thicklines
\path(5246,449)(5121,361)
\thinlines
\put(2576,1615){\blacken\ellipse{336}{336}}
\put(2576,1615){\ellipse{336}{336}}
\put(4256,1663){\blacken\ellipse{336}{336}}
\put(4256,1663){\ellipse{336}{336}}
\put(4656,3247){\blacken\ellipse{336}{336}}
\put(4656,3247){\ellipse{336}{336}}
\put(3296,4143){\blacken\ellipse{336}{336}}
\put(3296,4143){\ellipse{336}{336}}
\put(176,3807){\blacken\ellipse{336}{336}}
\put(176,3807){\ellipse{336}{336}}
\put(1440,255){\blacken\ellipse{336}{336}}
\put(1440,255){\ellipse{336}{336}}
\put(5456,175){\blacken\ellipse{336}{336}}
\put(5456,175){\ellipse{336}{336}}
\put(6496,3871){\blacken\ellipse{336}{336}}
\put(6496,3871){\ellipse{336}{336}}
\put(3296,6111){\blacken\ellipse{336}{336}}
\put(3296,6111){\ellipse{336}{336}}
\thicklines
\path(1708,161)(5371,161)(6471,3874)
	(3271,6136)(196,3786)(1458,236)
	(2571,1661)(4758,3249)(2021,3174)
	(4321,1599)(3296,4224)(2621,1886)
\path(2521,1924)(2721,1874)
\path(1708,236)(1708,74)
\path(446,3724)(1808,3236)
\path(3296,5874)(3296,4436)
\path(4921,3324)(6208,3786)
\path(4408,1449)(5196,386)
\path(483,3836)(433,3636)
\path(1846,3324)(1783,3174)
\path(3208,4424)(3383,4424)
\path(3208,5874)(3383,5874)
\path(6183,3886)(6258,3711)
\path(4871,3436)(4958,3236)
\path(4483,1486)(4358,1399)
\thinlines
\put(2112,3135){\blacken\ellipse{336}{336}}
\put(2112,3135){\ellipse{336}{336}}
\end{picture}
}}
\subfigure[]{\label{Figure:Petersen3}\setlength{\unitlength}{0.00021872in}
\begingroup\makeatletter\ifx\SetFigFont\undefined
\def\x#1#2#3#4#5#6#7\relax{\def\x{#1#2#3#4#5#6}}%
\expandafter\x\fmtname xxxxxx\relax \def\y{splain}%
\ifx\x\y   
\gdef\SetFigFont#1#2#3{%
  \ifnum #1<17\tiny\else \ifnum #1<20\small\else
  \ifnum #1<24\normalsize\else \ifnum #1<29\large\else
  \ifnum #1<34\Large\else \ifnum #1<41\LARGE\else
     \huge\fi\fi\fi\fi\fi\fi
  \csname #3\endcsname}%
\else
\gdef\SetFigFont#1#2#3{\begingroup
  \count@#1\relax \ifnum 25<\count@\count@25\fi
  \def\x{\endgroup\@setsize\SetFigFont{#2pt}}%
  \expandafter\x
    \csname \romannumeral\the\count@ pt\expandafter\endcsname
    \csname @\romannumeral\the\count@ pt\endcsname
  \csname #3\endcsname}%
\fi
\fi\endgroup
{\renewcommand{\dashlinestretch}{30}
\begin{picture}(6576,6881)(0,-10)
\put(3309,6530){\makebox(0,0)[lb]{\smash{{{\SetFigFont{5}{6.0}{rm}}}}}}
\put(5280,175){\blacken\ellipse{336}{336}}
\put(5280,175){\ellipse{336}{336}}
\put(4304,1631){\blacken\ellipse{336}{336}}
\put(4304,1631){\ellipse{336}{336}}
\put(2656,1711){\blacken\ellipse{336}{336}}
\put(2656,1711){\ellipse{336}{336}}
\put(1968,3215){\blacken\ellipse{336}{336}}
\put(1968,3215){\ellipse{336}{336}}
\put(176,3807){\blacken\ellipse{336}{336}}
\put(176,3807){\ellipse{336}{336}}
\put(3312,4239){\blacken\ellipse{336}{336}}
\put(3312,4239){\ellipse{336}{336}}
\put(4752,3183){\blacken\ellipse{336}{336}}
\put(4752,3183){\ellipse{336}{336}}
\put(6400,3919){\blacken\ellipse{336}{336}}
\put(6400,3919){\ellipse{336}{336}}
\put(3296,6111){\blacken\ellipse{336}{336}}
\put(3296,6111){\ellipse{336}{336}}
\thicklines
\path(3496,6018)(6434,3880)(4709,3193)
	(2634,1643)(3234,3955)
\path(134,3493)(409,3593)
\path(3109,3993)(3321,3930)
\path(3559,6093)(3434,5905)
\path(2509,1443)(1621,330)
\path(1546,405)(1671,280)
\path(2396,1493)(2571,1368)
\path(6396,3655)(5434,393)
\path(5346,443)(5534,343)
\path(6296,3680)(6509,3618)
\path(2334,3230)(4509,3230)
\path(2321,3330)(2321,3180)(2321,3155)
\path(4484,3330)(4484,3118)
\path(271,3555)(1471,205)(5346,193)
	(4271,1668)(3321,4255)(3321,6168)
	(171,3793)(2021,3230)(4046,1805)
\path(4096,1893)(3971,1743)
\thinlines
\put(1472,175){\blacken\ellipse{336}{336}}
\put(1472,175){\ellipse{336}{336}}
\end{picture}
}}
\caption{Three compatible normal odd partitions of the Petersen's
graph} \label{Figure:3PNICdePetersen}
\end{figure}
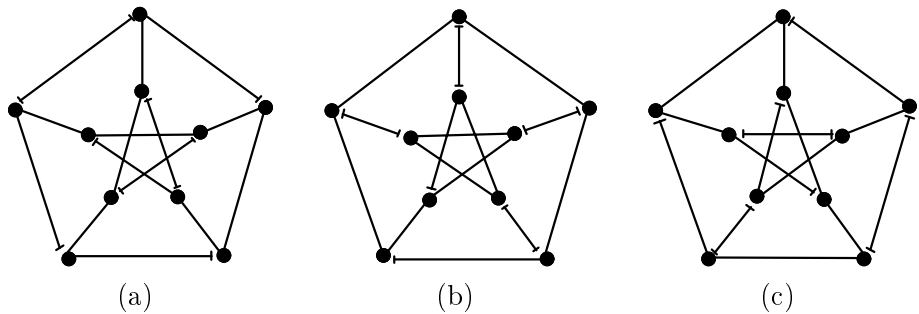
It seems difficult to characterize the class of cubic graphs with
three compatible normal odd partitions. The Petersen's graph has this
property (see Figure \ref{Figure:3PNICdePetersen}). In a forthcoming
paper we prove that $3$-edge colorable graphs also have this
property as well as the {\em flower snarks}.
\bibliographystyle{plain}

\bibliography{NormalPartition}

\end{document}